\documentclass[pra,twocolumn,superscriptaddress]{revtex4}
\usepackage{amsmath}
\usepackage{amsbsy}
\usepackage{footnote}
\usepackage{graphicx}
\usepackage{dsfont}
\usepackage{epstopdf}
\begin{document}

\title{Continuous Input Nonlocal Games}

\author{N. Aharon}
\author{S. Machnes}
\author{B. Reznik}
\author{J. Silman}
\author{L. Vaidman}
\affiliation{School of Physics and Astronomy, Tel-Aviv University, Tel-Aviv 69978, Israel}


\begin{abstract}
\textbf{We present a family of nonlocal games in which the inputs the players receive are continuous. We study three representative members of the family. For the first two a team sharing quantum correlations (entanglement) has an advantage over any team restricted to classical correlations. We conjecture that this is true for the third member of the family as well.}
\end{abstract}

\maketitle

The nonlocal nature of quantum mechanics, as manifested in
Bell inequalities violation  \citep{Bell,CHSH}, has recently been
highlighted in a number of games \citep{Tsirelson,VaidmanGHZ,Vaidmannecklace,Cabello,Aravind,Cleve,Brassard}.
Termed nonlocal \citep{Cleve}, these are cooperative games with incomplete
information for a team of remote players. Each of the players is
assigned by a verifier an input generated according to a known joint
probability distribution. The players must then send an output to
the verifier, who carries a truth table dictating for each
combination of inputs, which combinations of outputs result in a
win. The players may coordinate a joint strategy prior to receiving
their input, but cannot communicate with one another subsequently. A
team sharing quantum correlations (entanglement) is
said to employ a ``quantum strategy," while a team restricted to
sharing classical correlations is said to employ a ``classical
strategy."\\

In this paper we analyze three representative members of a novel family of nonlocal games, which differ from
other nonlocal games in the literature in that the input sets are continuous rather than discrete
and finite.
Moreover, most nonlocal games include a ``promise'' regarding the allowed input combinations and their frequency.
This means that the joint probability distribution governing the assignment of combinations of inputs is not uniform.
This restriction is especially tailored to guarantee a maximum
quantum advantage, and can make the rules of the game complex. In the games that we analyze
there is no such promise. The joint probability distribution governing
the assignment of inputs is uniform, and the rules are simple. Nevertheless, a non-negligible
quantum advantage obtains.\\

In the first game two remote players $A$ and $B$ receive a uniformly
generated input $a\in[0,\,1]$ and $b\in[0,\,1]$, respectively. Following this,
 each of the players sends a classical bit
representing an output $o_{i}\in\{1,\,-1\}$ ($i=A,\, B$) to the
verifier. The game is considered to have been won if \begin{equation}
o_{A}\cdot o_{B}=\left\{ \begin{array}{cc}
+1\,, & a+b<1\\
\\
-1\,, & a+b\geq1\end{array}\right..\label{conditions}\end{equation}
The game, therefore, amounts to the problem of returning a positive (negative) product of outputs when the
sum of the inputs is less than (greater than or equal to) $1$. In the following we show that
a team employing a quantum strategy can
achieve a higher probability for winning the game than a team restricted
to classical strategies.

We begin by presenting the optimal classical strategy. It is easy to
show that it is deterministic, i.e. the output is a single-valued function of the
input, and is given for example by \begin{equation}
o_{A}=1\,,\qquad\qquad o_{B}=\left\{ \begin{array}{cc}
+1\,, & b<\frac{1}{2}\\
\\
-1\,, & b\geq\frac{1}{2}\end{array}\right..\label{g1 cla str}\end{equation}
The winning probability then equals $75\%$ (see Fig. 1). This may be verified by noting that the game can be cast as the continuum limit of a family of Bell inequalities, first discovered by Gisin \citep{Gisin},  for which Tsirelson proved  both the classical and quantum bounds \citep{Tsirelson2}. For more details see  \citep{SMA}.
\begin{figure}
\centering
\includegraphics[scale=0.45]{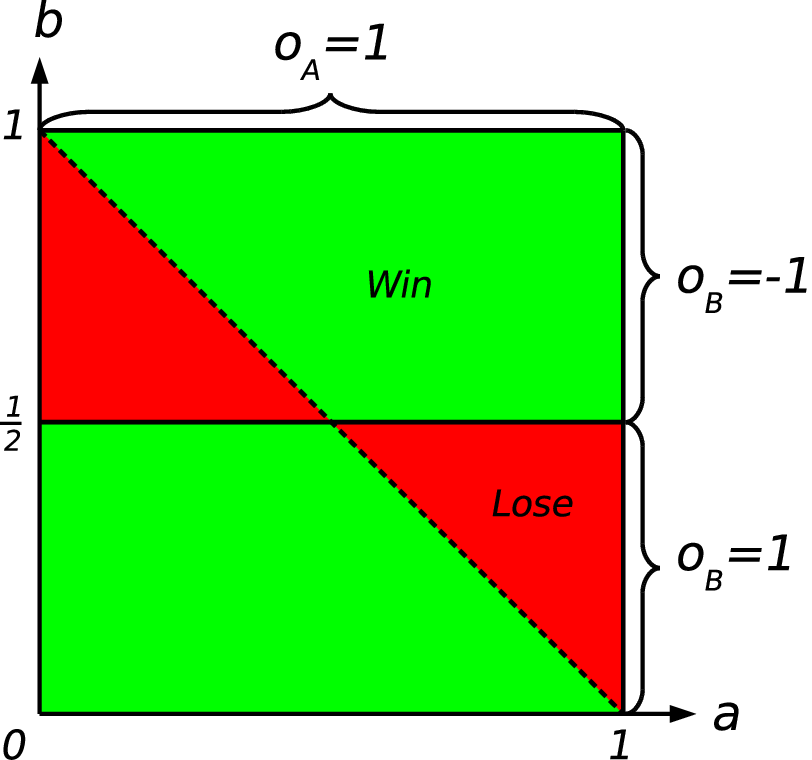}
\caption{Game 1 - the classical strategy. The lower (upper) big triangle is the region where identical  (opposite) outputs are required to win. Given the choice of outputs regions in which the game is won (lost) are colored in green (red). It is easy to see that the green regions add up to $\frac{3}{4}$ of the total area of the square.}
\end{figure}

In the quantum strategy we present the players share a two qubit singlet
state \begin{equation}
\left|\psi_{s}\right\rangle =\frac{1}{\sqrt{2}}(\left|\uparrow\downarrow\right\rangle -\left|\downarrow\uparrow\right\rangle )\,.\label{singlet}\end{equation}
Having beforehand agreed on a coordinate system, the players then
measure the spin component of their qubits along different axes in
the $xy$-plane. The choice of axes is dictated by the inputs as follows:
$A$ measures along an axis spanning an angle of $\theta_{A}(a)$
from the negative $x$-axis, while $B$ measures along an axis spanning
an angle of $\theta_{B}(b)$ from the negative $y$-axis (see Fig.
2).
\begin{figure}
\centering
\includegraphics[scale=0.50]{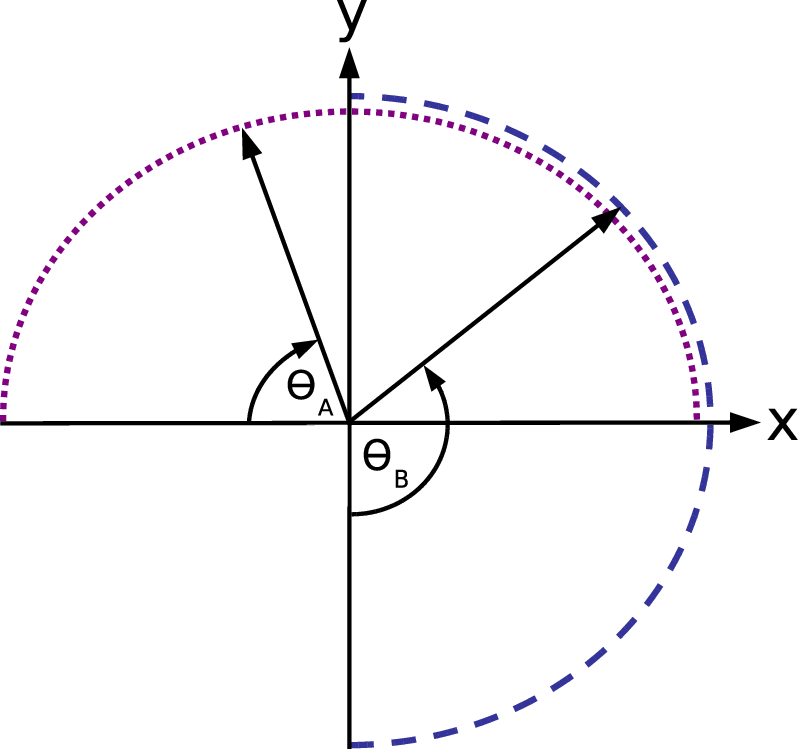}

\caption{Game 1 - the quantum strategy. $\theta_{A}$ and $\theta_{B}$ denote the
angles at which players $A$ and $B$, respectively, measure the spin
of their qubit. The dotted and dashed arcs denote the range of $\theta_{A}$
and $\theta_{B}$.}
\end{figure}
The players then send the results of the measurements to the verifier.
For $a+b\geq1$ the game is won if the two players obtain opposite
results, while for $a+b<1$ the converse holds. Given $a$ and $b$
the probability for identical results is
$\sin^{2}(\frac{\Delta}{2})$, where $\Delta\equiv\frac{3\pi}{2}-\theta_{A}(a)-\theta_{B}(b)$
is the angle between the axes of measurement. The winning probability
 is therefore given by \begin{eqnarray}
P_{W}  & = &  \int_{0}^{1}da\int_{0}^{1}db[\Theta(a+b-1)\cos^{2}(\frac{\Delta}{2}) \nonumber\\
  & & +\Theta(1-a-b)\sin^{2}(\frac{\Delta}{2})]\,,\label{prob game 1}\end{eqnarray}
where $\Theta$ is the unit step function $(\Theta(0)=1)$. To maximize $P_{W}$
we look for $\theta_{A}(a)$ and $\theta_{B}(b)$ such that when $a+b\geq1$
($a+b<1$) $\Delta$ is small (large). A most natural choice is \begin{equation}
\theta_{A}(a)=\pi a\,,\qquad\theta_{B}(b)=\pi b\,,\label{g1 ang func}\end{equation}
as is evident from Fig. 2. The integral then equals $\frac{1}{2}+\frac{1}{\pi}$ corresponding to a winning probability of
$\approx81.8\%$ and saturating the Tsirelson bound of the corresponding Bell inequality \citep{Tsirelson2}. This gives an advantage of $\approx6.8\%$ to a team making
use of quantum correlations over a team limited to classical correlations.\\

The above game is a special case of a more general joint task in which $A$ and $B$ are assigned the uniformly generated inputs $a\in[0,\,m]$ and $b\in[0,\,n]$, respectively, and must return correlated (anticorrelated) outputs when $a+b<\frac{n+m}{2}$  ($a+b\geq\frac{n+m}{2}$). Note that by setting $n=-m$ and defining $\tilde{b}\equiv -b$, the task reduces to having to return identical outputs when $a<\tilde{b}$ and
opposite otherwise.\\

The second game is identical to the first in all but the winning conditions.
The game is now considered to have been won if \begin{equation}
o_{A}\cdot o_{B}=\left\{ \begin{array}{cc}
+1\,, & 4|b-a|\;\mathrm{mod}\,3>1  \\
\\
-1\,, & 4|b-a|\;\mathrm{mod}\,3\leq 1 \end{array}\right..\label{g2 conditions}\end{equation}
That is, the players must return correlated outputs if the absolute value of the their inputs' difference is in the interval $\left[\frac{1}{4},\,\frac{3}{4}\right]$, otherwise they must return anticorrelated outputs.

A possible realization of the optimal classical strategy is \begin{equation}
o_{A}=\left\{ \begin{array}{cc}
+1\,, & a\leq\frac{1}{2}\\
\\
-1\,, & a>\frac{1}{2}\end{array}\right.,\qquad o_{B}=\left\{ \begin{array}{cc}
-1\,, & b\leq\frac{1}{2}\\
\\
+1\,, & b>\frac{1}{2}\end{array}\right..\label{g2 cla str}\end{equation}
The winning probability equals $75\%$, as in the first game (see Fig. 3). To see that this is the maximum, consider Fig. 3. If we cyclically shift the input of one of the players by $\frac{1}{4}$, then the regions that require correlated or anticorrelated outputs within each quadrant correspond to the first game \citep{note1}. Therefore, if the game admitted a strategy with a winning probability greater than $75\%$ in any of the quadrants, so would the first game.
\begin{figure}
\centering
\includegraphics[scale=0.45]{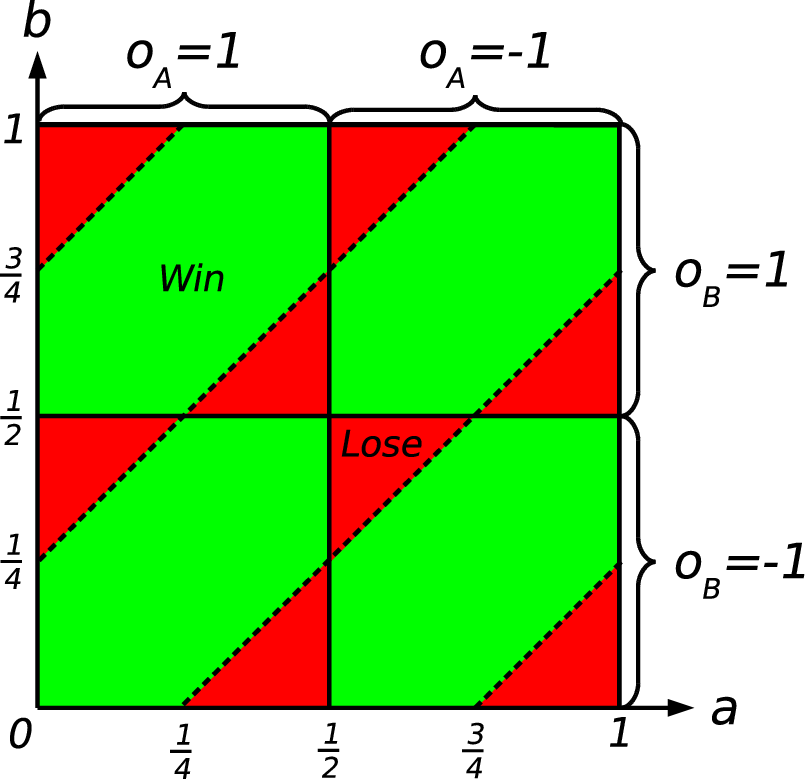}

\caption{Game 2 - the classical strategy. The two small triangles and the strip between the two middle dashed lines are regions where identical outputs are required to win. Given the choice of outputs regions where the game is won (lost) regions are colored in green (red). The green regions add up to $\frac{3}{4}$ of the total area of the square.}
\end{figure}
The quantum strategy we present differs from that of the first game only
in the choice of axes $A$ and $B$ measure along. The winning probability now equals

\begin{eqnarray}
P_{W}  & = & \int_{0}^{1}da\int_{0}^{1}db[\Theta(4|b-a|\;\mathrm{mod}\,3-1)\cos^{2}(\frac{\Delta}{2}) \nonumber \\
 & & +\Theta(1-4|b-a|\;\mathrm{mod}\,3)\sin^{2}(\frac{\Delta}{2})]\,,\label{prob game 2}\end{eqnarray}
Here $\Delta\equiv\theta_{A}(a)-\theta_{B}(b)$ with both angles now spanning from the $y$-axis in the $xy$-plane. The maximum obtains for  \begin{equation}
\theta_{A}(a)=2\pi a\,,\qquad\theta_{B}(b)=2\pi b\,,\label{g2 qua str}\end{equation}
giving the same winning probability as in the first game, i.e. $\approx81.8\%$, and equalling the Tsirelson bound of the corresponding Bell inequality \citep{Tsirelson2}.

Both games described naturally accommodate a geometric description. For
example, as is evident from the quantum strategy, the second game
can be reformulated as the problem of returning identical outputs when the angle between
a pair of nonvanishing two-dimensional vectors is greater than $\frac{\pi}{2}$.
The question arises as to how the quantum advantage changes when playing the game in three dimensions.
 More specifically, two remote
players are each assigned a pair of angles $0\leq\theta_{i}\leq\pi$,
$0\leq\varphi_{i}<2\pi$, designating a three dimensional unit vector
$\hat{\boldsymbol{\mathrm{r}}}_{i}$ ($i=A,\, B$). The game is considered to have been won if
\begin{equation}
o_{A}\cdot o_{B}=\left\{ \begin{array}{cc}
+1\,, & \hat{\boldsymbol{\mathrm{r}}}_{A}\cdot\hat{\boldsymbol{\mathrm{r}}}_{B}<0\\
\\
-1\,, & \hat{\boldsymbol{\mathrm{r}}}_{A}\cdot\hat{\boldsymbol{\mathrm{r}}}_{B}\geq0\end{array}\right..\label{g3 conditions}\end{equation}
\\
The joint probability distribution governing the assignment of angles
is a product $\rho_{A}(\theta_{A},\,\varphi_{A})\cdot\rho_{B}(\theta_{B},\,\varphi_{B})$
with \begin{equation}
\rho_{i}(\theta_{i},\,\varphi_{i})=\sin\theta_{i}\,,\label{dist}\end{equation}
guaranteeing isotropy \citep{note2}.
The classical strategy that we present is an extension of the optimal classical strategy of the second game, where in the geometric description  $A$ ($B$) returns an output
equal to $1$ ($-1$), respectively, if the angle corresponding to his input is less than or equal to $\pi$. Otherwise, $A$ ($B$) returns $-1$ ($1$). Similarly, we now have $A$ ($B$) return $1$ ($-1$) when $\theta_{A}\leq\frac{\pi}{2}$ ($\theta_{B}\leq\frac{\pi}{2}$), independent of $\varphi_{A}$ ($\varphi_{B}$), and $-1$ ($1$) otherwise. This gives  $\approx68.2\%$ ($1-\frac{1}{\pi}$) probability of winning. It seems likely that this strategy is the optimal.

As in the other games, in the quantum strategy that we consider, $A$ and $B$ share a singlet state of two qubits and measure along axes dictated by their inputs, $\hat{\boldsymbol{\mathrm{n}}}_{A}(\hat{\boldsymbol{\mathrm{r}}}_{A})$ and $\hat{\boldsymbol{\mathrm{n}}}_{B}(\hat{\boldsymbol{\mathrm{r}}}_{B})$. The probability for winning is then given by
\begin{eqnarray}
P_{W} & = & \int_{\Omega_{A}}d\Omega_{A}\int_{\Omega_{B}}d\Omega_{B}[\Theta(\hat{\boldsymbol{\mathrm{r}}}_{A}\cdot\hat{\boldsymbol{\mathrm{r}}}_{B})\cos^{2}(\frac{\Delta}{2})\nonumber\\ &  &+\Theta(-\hat{\boldsymbol{\mathrm{r}}}_{A}\cdot\hat{\boldsymbol{\mathrm{r}}}_{B})\sin^{2}(\frac{\Delta}{2})]\,,\label{prob game 3}\end{eqnarray}
\\
with $\Delta\equiv \arccos(\hat{\boldsymbol{\mathrm{n}}}_{A}(\hat{\boldsymbol{\mathrm{r}}}_{A})\cdot\hat{\boldsymbol{\mathrm{n}}}_{B}(\hat{\boldsymbol{\mathrm{r}}}_{B}))$, and maximizes for \begin{equation}
\hat{\boldsymbol{\mathrm{n}}}_{A}(\hat{\boldsymbol{\mathrm{r}}}_{A})=\hat{\boldsymbol{\mathrm{r}}}_{A}\,,\qquad\hat{\boldsymbol{\mathrm{n}}}_{B}(\hat{\boldsymbol{\mathrm{r}}}_{B})=\hat{\boldsymbol{\mathrm{r}}}_{B}\,.\label{g3 qua str}\end{equation} The probability of winning than equals $75\%$. Numerical evidence obtained using semi-definite programming (SDP) indicates that this stratgey is optimal. Interestingly, the quantum advantage remains unchanged equaling $\approx6.8\%$.\\

In fact, all the games share a unifying  ``theme". Suppose that $A$ and $B$ each receive the coordinates of a randomly generated three dimensional vector  ${\boldsymbol{\mathrm{r}}}_{A}$ and ${\boldsymbol{\mathrm{r}}}_{B}$, respectively. Then by a suitable choice of the joint probability distribution governing the assignment of the vectors, each of the games translates to a question about the quantity

\begin{equation}
\xi\equiv |\boldsymbol{\mathrm{r}}_{B}-\boldsymbol{\mathrm{r}}_{A}|=\sqrt{\boldsymbol{\mathrm{r}}_{B}^{2}-2{\boldsymbol{\mathrm{r}}}_{B}\cdot {\boldsymbol{\mathrm{r}}}_{A}+\boldsymbol{\mathrm{r}}_{A}^{2}}\,.\label{unify}\end{equation}
The third game obtains if we restrict the vectors to unit magnitude. Actually, it is enough to require that the vectors be nonvanishing so long as they are generated isotropically. We then ask whether $\xi<\sqrt{\boldsymbol{\mathrm{r}}_{B}^{2}+\boldsymbol{\mathrm{r}}_{A}^{2}}$. The second game is identical except that we further restrict the vectors to lie on the same plane. In the first game we abolish isotropy altogether. The vectors are generated anitparallel to one another, with their magnitudes uniformly distributed between $0$ and $1$. $\xi$ then equals $r_{A}+r_{B}$, and the players must decide whether $\xi>1$. In particular, we see that by asking different questions and imposing different  constraints we obtain different games. In this sense the three games can be considered as belonging to a larger family of games.\\

\textbf{Acknowledgments} We acknowledge support from the Israeli Science Foundation (grants no. 784/06 and 990/06), and from the European Commission under the Integrated Project Qubit Applications (QAP) funded by the IST Directorate (contract no. 015848).


\begin{thebibliography}{45}
\bibitem{note1}To be more precise, each of the quadrants corresponds to the truth table of the $a<b$ or $a>b$ formulation of the first game.
\bibitem{note2}The differential of a solid angle, $\Omega$, in spherical coordinates is proportional to $\sin\theta$. This introduces a weight function when integrating over $\theta$ and $\varphi$.
\end{thebibliography}
\end{document}